\documentstyle[aps,12pt]{revtex}
\begin{document}
\title{Relativistic Hartree-Bogoliubov description\\
of the deformed ground-state proton emitters}
\author{D. Vretenar$^{1,2}$, G.A. Lalazissis$^{2}$, and P. Ring$^{2}$}
\address{
$^{1}$ Physics Department, Faculty of Science, University of
Zagreb, Croatia\\
$^{2}$ Physik-Department der Technischen Universit\"at M\"unchen,
D-85748 Garching, Germany\\}
\maketitle
\bigskip
\bigskip
\begin{abstract}
Ground-state properties of deformed proton-rich odd-Z
nuclei in the region $59 \leq Z \leq 69$ are described in the framework
of Relativistic Hartree Bogoliubov (RHB) theory. 
One-proton separation energies and ground-state quadrupole
deformations that result from fully self-consistent microscopic 
calculations are compared with available experimental data.
The model predicts the location of the 
proton drip-line, the properties of proton emitters beyond the drip-line, 
and provides information about the deformed single-particle 
orbitals occupied by the odd valence proton.
\end{abstract}
\vspace{1 cm}
{PACS numbers:} {21.60.Jz, 21.10.Dr, 23.50.+z, 27.60+j}\\
\vspace{1 cm}\\
The decay by direct proton emission 
provides the opportunity to study the structure of systems beyond the 
drip-line. The phenomenon of ground-state proton radioactivity
is determined by a delicate interplay between the Coulomb and
centrifugal terms of the effective potential. While low-Z 
nuclei lying beyond the proton drip-line exist only as short 
lived resonances, the relatively high potential energy barrier 
enables the observation of ground-state proton emission 
from medium-heavy and heavy nuclei. At the drip-lines proton 
emission competes with $\beta^+$ decay; for heavy nuclei 
also fission or $\alpha$ decay can be favored. The proton 
drip-line has been fully mapped up to $Z=21$, and possibly for
odd-Z nuclei up to In \cite{WD.97}. No examples of ground-state
proton emission have been discovered below $Z=50$. 
Proton radioactivity has been studied from odd-Z nuclei
in the two spherical regions from $51\leq Z \leq 55$
and $69\leq Z \leq 83$. The systematics of proton decay 
spectroscopic factors is consistent with half-lives calculated 
in the spherical WKB or DWBA approximations. 
Recently reported proton decay rates \cite{Dav.98} indicate
that the missing region of light rare-earth nuclei contains strongly deformed
nuclei at the drip-lines. The lifetimes of deformed proton emitters
provide direct information on the last occupied Nilsson configuration,
and therefore on the shape of the nucleus. Modern models for proton
decay rates from deformed nuclei have only recently been developed
\cite{MFL.98}. However, even the most realistic calculations are not 
based on a fully microscopic and self-consistent description 
of proton unstable nuclei. In particular, such a description 
should also include important pairing correlations.

The model that we use for the description of ground-state properties 
of proton emitters, is formulated in the framework of
Relativistic Hartree Bogoliubov (RHB) theory. Models of 
quantum hadrodynamics that are
based on the relativistic mean-field approximation, have
been successfully applied in the description of a variety of nuclear
structure phenomena in spherical and deformed $\beta$-stable nuclei \cite
{Rin.96}, and more recently in studies of exotic nuclei far from the
valley of beta stability. 
RHB presents a relativistic extension of the Hartree-Fock-Bogoliubov
theory. It was derived in Ref. \cite{KR1.91} in an attempt to
develop a unified framework in which  relativistic mean-field and
pairing correlations could be described simultaneously. Such a 
unified and self-consistent formulation is especially important
in applications to drip-line nuclei, in which the separation
energy of the last nucleons can become extremely small, the Fermi level
is found close to the particle continuum, and the
lowest particle-hole or particle-particle modes are embedded in the
continuum. The RHB model with finite range pairing interactions has been 
used to describe halo phenomena in light nuclei \cite{PVL.97}, 
the neutron drip-line in light nuclei \cite{LVP.98}, the 
reduction of the spin-orbit potential in drip-line nuclei \cite{LVR.97},
ground-state properties of Ni and Sn isotopes \cite{LVR.98}, 
the deformations and shape coexistence phenomena that result 
from the suppression of the spherical $N = 28$ shell gap in 
neutron-rich nuclei \cite{LVR.98a}. In particular, in Ref. \cite{VLR.98}
we have applied the RHB model to describe 
properties of proton-rich spherical even-even nuclei
$14\leq Z \leq 28$ and $N=18,20,22$. It has been shown that the 
model correctly predicts the location of the proton drip-line.
The isospin dependence of the effective spin-orbit potential
has been studied.

In the relativistic Hartree-Bogoliubov model, 
the ground state of a nucleus $\vert \Phi >$ is represented 
by the product of independent single-quasiparticle states.
These states are eigenvectors of the
generalized single-nucleon Hamiltonian which
contains two average potentials: the self-consistent mean-field
$\hat\Gamma$ which encloses all the long range particle-hole ({\it ph})
correlations, and a pairing field $\hat\Delta$ which sums
up the particle-particle ({\it pp}) correlations. The 
single-quasiparticle equations result from the variation of the 
energy functional with respect to the hermitian density matrix $\rho$
and the antisymmetric pairing tensor $\kappa$. 
In the Hartree approximation for
the self-consistent mean field, the relativistic
Hartree-Bogoliubov (RHB) equations read
\begin{eqnarray}
\label{equ.2.2}
\left( \matrix{ \hat h_D -m- \lambda & \hat\Delta \cr
                -\hat\Delta^* & -\hat h_D + m +\lambda} \right) 
\left( \matrix{ U_k({\bf r}) \cr V_k({\bf r}) } \right) =
E_k\left( \matrix{ U_k({\bf r}) \cr V_k({\bf r}) } \right).
\end{eqnarray}
where $\hat h_D$ is the single-nucleon Dirac
Hamiltonian \cite{Rin.96}, and $m$ is the nucleon mass
\begin{equation}
\hat{h}_{D}=-i{\mathbf{\alpha \cdot \nabla }}+\beta (m+g_{\sigma }\sigma
({\mathbf r}))+g_{\omega }\tau _{3}\omega ^{0}({\mathbf r})+
g_{\rho }\rho ^{0}({\mathbf r})+
e{\frac{{(1-\tau _{3})}}{2}}A^{0}({\mathbf r}).  \label{dirh}
\end{equation}
The chemical potential $\lambda$  has to be determined by
the particle number subsidiary condition in order that the
expectation value of the particle number operator
in the ground state equals the number of nucleons. The column
vectors denote the quasi-particle spinors and $E_k$
are the quasi-particle energies. The Dirac Hamiltonian 
contains the mean-field potentials of the isoscalar 
scalar $\sigma$-meson, the isoscalar vector $\omega$-meson,
the isovector vector $\rho$-meson, as well as the 
electrostatic potential. 
The RHB equations have to be solved self-consistently with
potentials determined in the mean-field approximation from
solutions of Klein-Gordon equations.
The equation for the sigma meson contains the 
non-linear $\sigma$ self-interaction terms.
Because of charge conservation only the
third component of the isovector $\rho$-meson contributes. The
source terms for the Klein-Gordon equations are calculated 
in the {\it no-sea} approximation. In the present version of the 
model we do not perform angular momentum or particle number
projection.

The pairing field $\hat\Delta $ in (\ref{equ.2.2}) is an integral
operator with the kernel 
\begin{equation}
\label{equ.2.5}
\Delta_{ab} ({\bf r}, {\bf r}') = {1\over 2}\sum\limits_{c,d}
V_{abcd}({\bf r},{\bf r}') {\bf\kappa}_{cd}({\bf r},{\bf r}'),
\end{equation}
where $a,b,c,d$ denote quantum numbers
that specify the Dirac indices of the spinors, 
$V_{abcd}({\bf r},{\bf r}')$ are matrix elements of a
general relativistic two-body pairing interaction, and the pairing
tensor is defined 
\begin{equation}
{\bf\kappa}_{cd}({\bf r},{\bf r}') = 
\sum_{E_k>0} U_{ck}^*({\bf r})V_{dk}({\bf r}').
\end{equation}
In most of the applications of the RHB model we have used
a phenomenological non relativistic interaction in the pairing 
channel: the pairing part of the Gogny force 
\begin{equation}
V^{pp}(1,2)~=~\sum_{i=1,2}
e^{-(( {\bf r}_1- {\bf r}_2)
/ {\mu_i} )^2}\,
(W_i~+~B_i P^\sigma 
-H_i P^\tau -
M_i P^\sigma P^\tau),
\end{equation}
with the set D1S \cite{BGG.84} for the parameters 
$\mu_i$, $W_i$, $B_i$, $H_i$ and $M_i$ $(i=1,2)$.  
This force has been very carefully adjusted to the pairing 
properties of finite nuclei all over the periodic table. 
In particular, the basic advantage of the Gogny force
is the finite range, which automatically guarantees a proper
cut-off in momentum space. 

The eigensolutions of Eq. (\ref{equ.2.2}) form a set of
orthogonal and normalized single quasi-particle states. The corresponding
eigenvalues are the single quasi-particle energies.
The self-consistent iteration procedure is performed
in the basis of quasi-particle states. 
A simple blocking procedure is used in the calculation of 
odd-proton and/or odd-neutron systems.
The resulting quasi-particle
eigenspectrum is then transformed into the canonical basis of single-particle
states, in which the RHB ground-state takes the  
BCS form. The transformation determines the energies
and occupation probabilities of the canonical states.

The input parameters are the coupling constants and
masses for the effective mean-field Lagrangian, and the
effective interaction in the pairing channel. As 
in most applications of the RHB model, 
we use the NL3 effective interaction~\cite{LKR.97}
for the RMF Lagrangian. 
Properties calculated with NL3 indicate that this is probably
the best RMF effective interaction so far, both for nuclei
at and away from the line of $\beta$-stability. For the pairing
field we employ the pairing part of the
Gogny interaction, with the parameter set D1S \cite{BGG.84}.

In Fig. \ref{figA} we display the one-proton separation 
energies
\begin{equation}
S_{p}(Z,N) = B(Z,N) - B(Z-1,N)
\label{sep}
\end{equation}
for the odd-Z nuclei $59 \leq Z \leq 69$, as function of the
number of neutrons. The model predicts the drip-line nuclei:
$^{124}$Pr, $^{129}$Pm, $^{134}$Eu, $^{139}$Tb, $^{146}$Ho, and
$^{152}$Tm. In heavy proton drip-line nuclei the potential energy 
barrier, which results from the superposition of the Coulomb and 
centrifugal potentials, is relatively high. For the proton 
decay to occur the odd valence proton must penetrate the 
potential barrier, and this process competes with $\beta^+$ decay.
Since the half-life for proton decay is inversely proportional 
to the energy of the odd proton, in many 
nuclei the decay will not be observed immediately after
the drip-line. Proton radioactivity is expected to dominate
over $\beta^+$ decay only when the energy of the odd proton 
becomes relatively high. This is also a crucial point for the
relativistic description of proton emitters, since the precise 
values of the separation energies depend on 
the isovector properties of the spin-orbit interaction.

The calculated separation energies should be compared with 
recently reported experimental data on proton radioactivity 
from $^{131}$Eu, $^{141}$Ho \cite{Dav.98}, $^{145}$Tm 
\cite{Bat.98}, and  $^{147}$Tm \cite{Sel.93}. The $^{131}$Eu
transition has an energy $E_p = 0.950 (8)$ MeV and a half-life
26(6) ms, consistent with decay from either $3/2^+[411]$ or
$5/2^+[413]$ Nilsson orbital. For $^{141}$Ho the transition 
energy is $E_p = 1.169 (8)$ MeV, and the half-life 4.2(4) ms 
is assigned to the decay of the $7/2^-[523]$ orbital.
The calculated proton separation energy, both for
$^{131}$Eu and $^{141}$Ho, is of $-0.9$ MeV. In 
the RHB calculation for $^{131}$Eu the
odd proton occupies the $5/2^+[413]$ orbital, while the 
ground state of $^{141}$Ho corresponds to the $7/2^-[523]$
proton orbital.  This orbital is also occupied by the odd
proton in the calculated ground states of $^{145}$Tm and
$^{147}$Tm. For the proton separation energies we obtain:
$-1.46$ MeV in $^{145}$Tm, and $-0.96$ MeV in $^{147}$Tm. These 
are compared with the experimental values for transition 
energies: $E_p = 1.728 (10)$ MeV in $^{145}$Tm, 
and $E_p = 1.054 (19)$ MeV in $^{147}$Tm. When compared 
with spherical WKB or DWBA calculations \cite{ASN.97}, the 
experimental half-lives for the two Tm isotopes are consistent
with spectroscopic factors for decays from the $h_{11/2}$ proton
orbital. Though our predicted ground-state configuration 
$7/2^-[523]$ indeed originates from the spherical 
$h_{11/2}$ orbital, we find that the two nuclei are  
deformed. $^{145}$Tm has a prolate quadrupole deformation
$\beta_2 = 0.23$, and $^{147}$Tm is oblate in the ground-state
with $\beta_2 = -0.19$. Calculations also predict possible
proton emitters $^{136}$Tb and $^{135}$Tb with separation
energies $-0.90$ MeV and $-1.15$ MeV, respectively. In both
isotopes the predicted ground-state proton configuration is 
$3/2^+[411]$.

The calculated mass quadrupole deformation parameters for 
the odd-Z nuclei $59 \leq Z \leq 69$ at and beyond the 
drip line are shown in Fig. \ref{figB}. Pr, Pm, Eu and
Tb isotopes are strongly prolate deformed 
($\beta_2 \approx 0.30 - 0.35$). By increasing the number 
of neutrons, Ho and Tm display a transition from prolate to
oblate shapes. The absolute values of $\beta_2$ decrease as
we approach the spherical solutions at $N = 82$. The quadrupole
deformations calculated in the RHB model with the NL3 
effective interaction, are found in excellent agreement with the
predictions of the macroscopic-microscopic mass model \cite{MN.95}.

A detailed analysis of single proton levels, including 
spectroscopic factors, can be performed in the canonical basis
which results from the fully microscopic and
self-consistent RHB calculations. For the Eu isotopes 
this is illustrated in Fig. \ref{figC}, where
we display the proton single-particle
energies in the canonical basis as function of the neutron number.
The thick dashed line denotes the position of the Fermi level.
The proton energies are the diagonal matrix elements of the
Dirac hamiltonian $h_D$ (\ref{dirh}) in the canonical basis. 
The phase-space of positive-energy states should not be confused
with the continuum of scattering states which asymptotically
behave as plane waves. The RHB ground-state wave function can be written
either in the quasiparticle basis as a product of
independent quasi-particle states, or in the {\it canonical basis}
as a highly correlated BCS-state. In the {\it canonical basis}
nucleons occupy  single-particle states. The canonical states
are eigenstates of the RHB density matrix. The eigenvalues are
the corresponding occupation numbers. In particular, we notice 
that for the proton emitter $^{131}$Eu, the ground-state corresponds
to the odd valence proton in the $5/2^+[413]$ orbital.

In conclusion, this study presents a detailed analysis of 
deformed proton emitters $59 \leq Z \leq 69$ in the 
framework of the relativistic Hartree-Bogoliubov theory. 
We have investigated the location of the proton drip-line, the    
separation energies for proton emitters beyond the 
drip-line, and ground-state quadrupole deformations.
The NL3 effective interaction has been used for the 
mean-field Lagrangian, and pairing correlations have been described 
by the pairing part of the finite range Gogny interaction D1S. 
The RHB results for proton separation energies are found to be in very good 
agreement with recent experimental data on direct proton decay 
of $^{131}$Eu, $^{141}$Ho, and $^{147}$Tm. The theoretical value
is not so good for $^{145}$Tm; the calculated and experimental
proton energies differ by more than 200 keV. Predictions for the 
deformed single-particle orbitals occupied by the valence odd protons, 
are consistent with experimental half-lives for proton
transitions. The model also predicts possible proton emitters
$^{136}$Tb and $^{135}$T. The present analysis has shown that the RHB  
with finite range pairing provides a fully self-consistent microscopic
model which can be used to map the entire proton drip-line for 
medium-heavy and heavy nuclei $51 \leq Z \leq 83$.
  
\bigskip
\begin{center}
{\bf ACKNOWLEDGMENTS}
\end{center}

This work has been supported in part by the 
Bundesministerium f\"ur Bildung und Forschung under
project 06 TM 875.

\newpage

\centerline{\bf Figure Captions}
\bigskip

\begin{figure}
\caption{ Calculated one-proton separation energies  
for odd-Z nuclei $59 \leq Z \leq 69$ at and beyond the 
drip-line.}
\label{figA}
\end{figure}

\begin{figure}
\caption{ Self-consistent ground-state quadrupole deformations   
for the odd-Z nuclei $59 \leq Z \leq 69$ at the proton drip-line.}
\label{figB}
\end{figure}

\begin{figure}
\caption{The proton single-particle levels
for the Eu isotopes. The dashed line denotes the position
of the Fermi level. The energies in the canonical basis
correspond to ground-state solutions calculated with the
NL3 effective force of the mean-field
Lagrangian. The parameter set D1S is used for
the finite range Gogny-type interaction in the
pairing channel.}
\label{figC}
\end{figure}


\begin{references}
\bibitem{WD.97} P.J. Woods and C.N. Davids, Annu. Rev. Nucl. Part.
	Sci. {\bf 47}, 541 (1997).
\bibitem{Dav.98} C.N. Davids {\it et al},
	Phys. Rev. Lett. {\bf 80}, 1849 (1998).
\bibitem{MFL.98} E. Maglione, L.S. Ferreira, and R.J. Liotta,
	Phys. Rev. Lett. {\bf 81}, 538 (1998).
\bibitem{Rin.96} P. Ring,  Progr. Part. Nucl. Phys. {\bf 37}, 193 (1996).
\bibitem{KR1.91} H. Kucharek and P. Ring; Z. Phys. {\bf A 339}, 23 (1991).
\bibitem{PVL.97} W. P\"oschl, D. Vretenar, G.A. Lalazissis,
        and P. Ring, Phys. Rev. Lett. {\bf 79}, 3841 (1997).
\bibitem{LVP.98} G.A. Lalazissis, D. Vretenar, W. P\"oschl,
        and P. Ring, Nucl. Phys. {\bf A632}, 363 (1998).
\bibitem{LVR.97} G.A. Lalazissis, D. Vretenar, W. P\"oschl,
        and P. Ring, Phys. Lett. {\bf B418}, 7 (1998).
\bibitem{LVR.98} G.A. Lalazissis, D. Vretenar,
        and P. Ring, Phys. Rev. C {\bf 57}, 2294 (1998).
\bibitem{LVR.98a} G.A. Lalazissis, D. Vretenar,
	and P. Ring, submitted to Phys. Rev. C
\bibitem{VLR.98} D. Vretenar,  G.A. Lalazissis, and P. Ring,
        Phys. Rev. C {\bf 57}, 3071 (1998).
\bibitem{BGG.84}  J. F. Berger, M. Girod and D. Gogny; Nucl. Phys. 
                 {\bf A428}, 32 (1984).
\bibitem{LKR.97}  G. A. Lalazissis, J. K\"{o}nig and P. Ring; Phys. Rev. 
        {\bf C55}, 540 (1997).
\bibitem{Bat.98} J.C. Batchelder {\it et al},
	Phys. Rev. C {\bf 57}, R1042 (1998).
\bibitem{Sel.93} P.J. Sellin {\it et al},
	Phys. Rev. C {\bf 47}, 1933 (1993).
\bibitem{ASN.97} S. \. Aberg, P.B. Semmes, and W. Nazarewicz,
	Phys. Rev. C {\bf 56}, 1762 (1997).
\bibitem{MN.95} P. M\" oller, J.R. Nix, W.D. Myers, and 
	W.J. Swiatecki, At. Data Nucl. Data Tables {\bf 59}, 185 (1995).
    
\end{references}
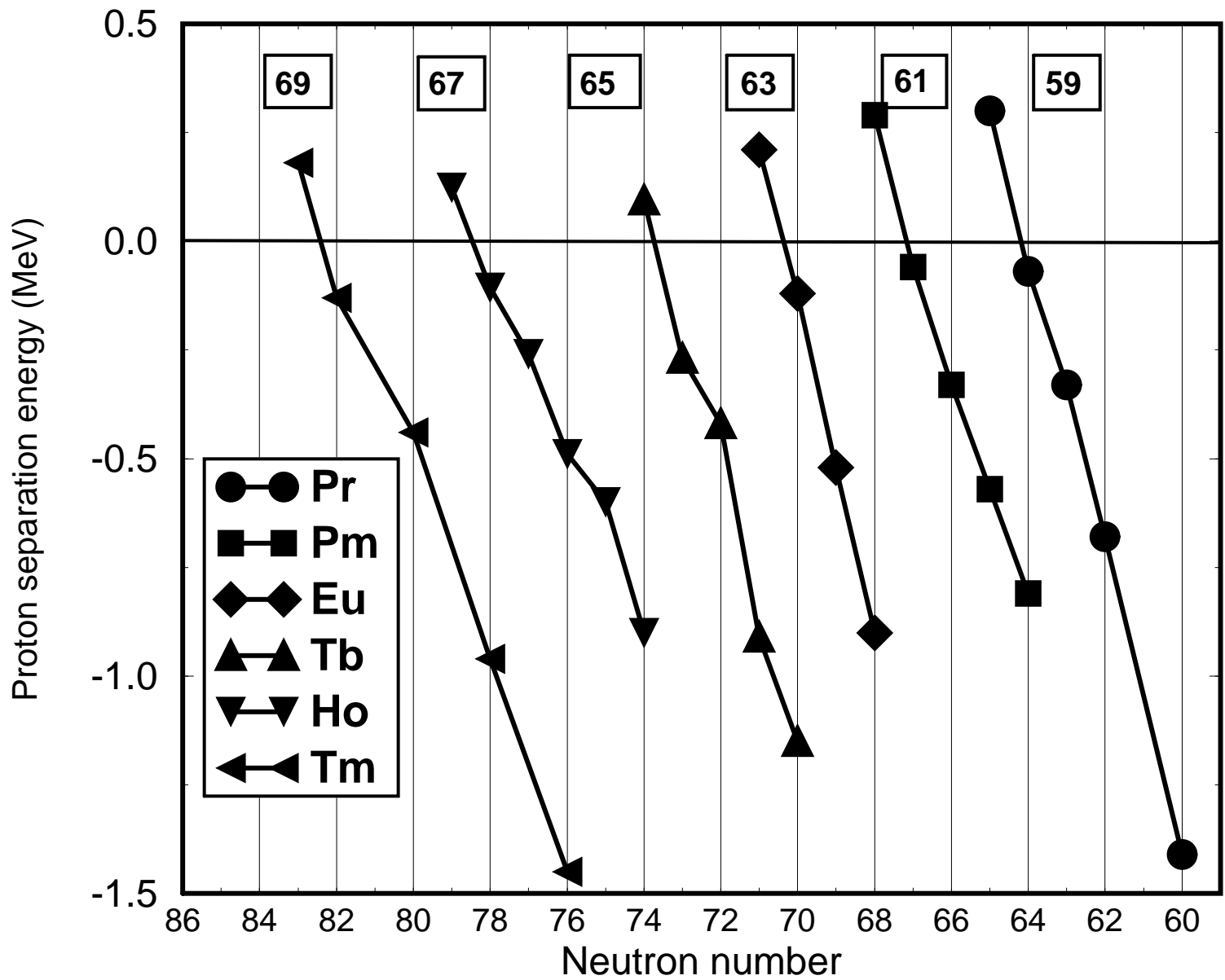
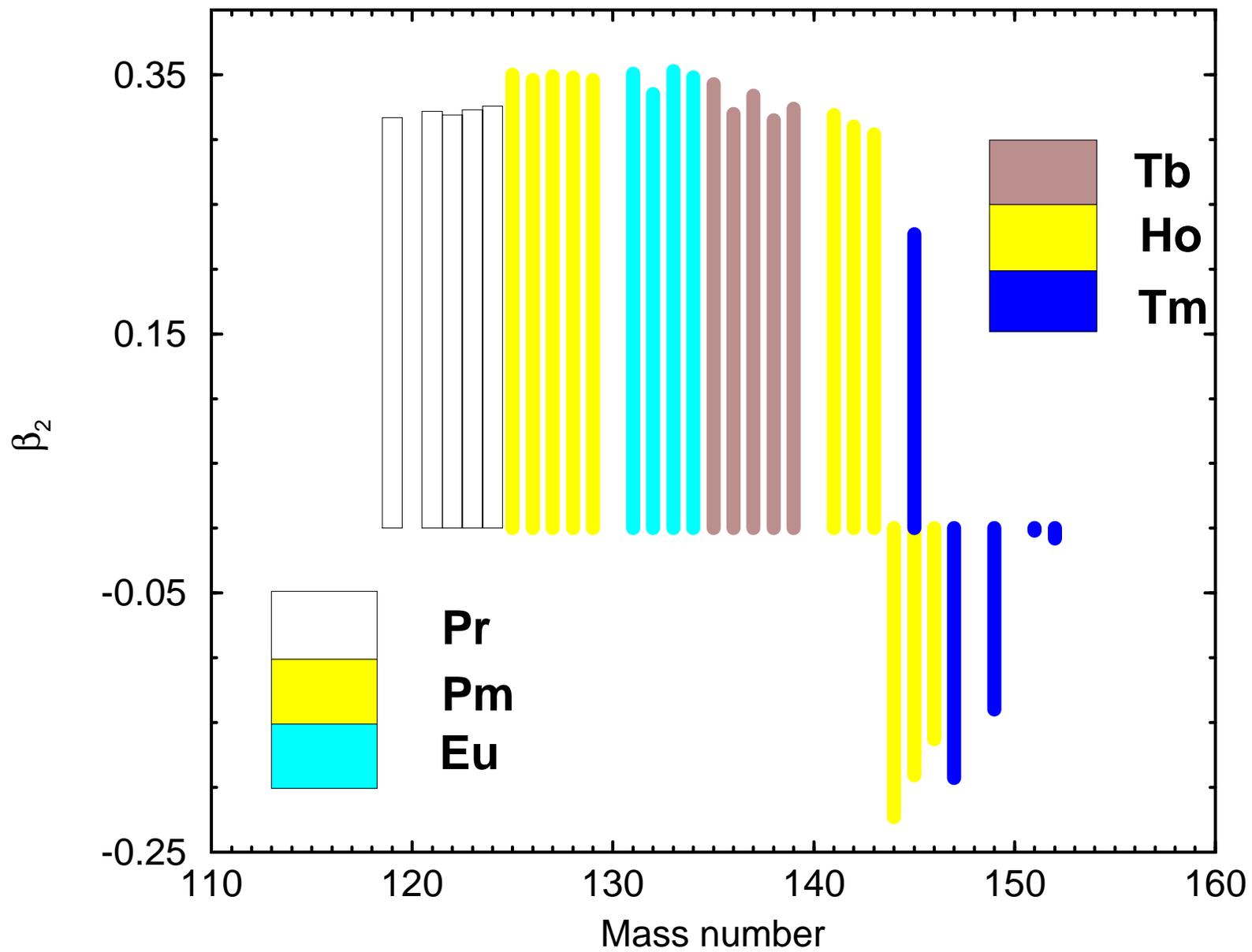
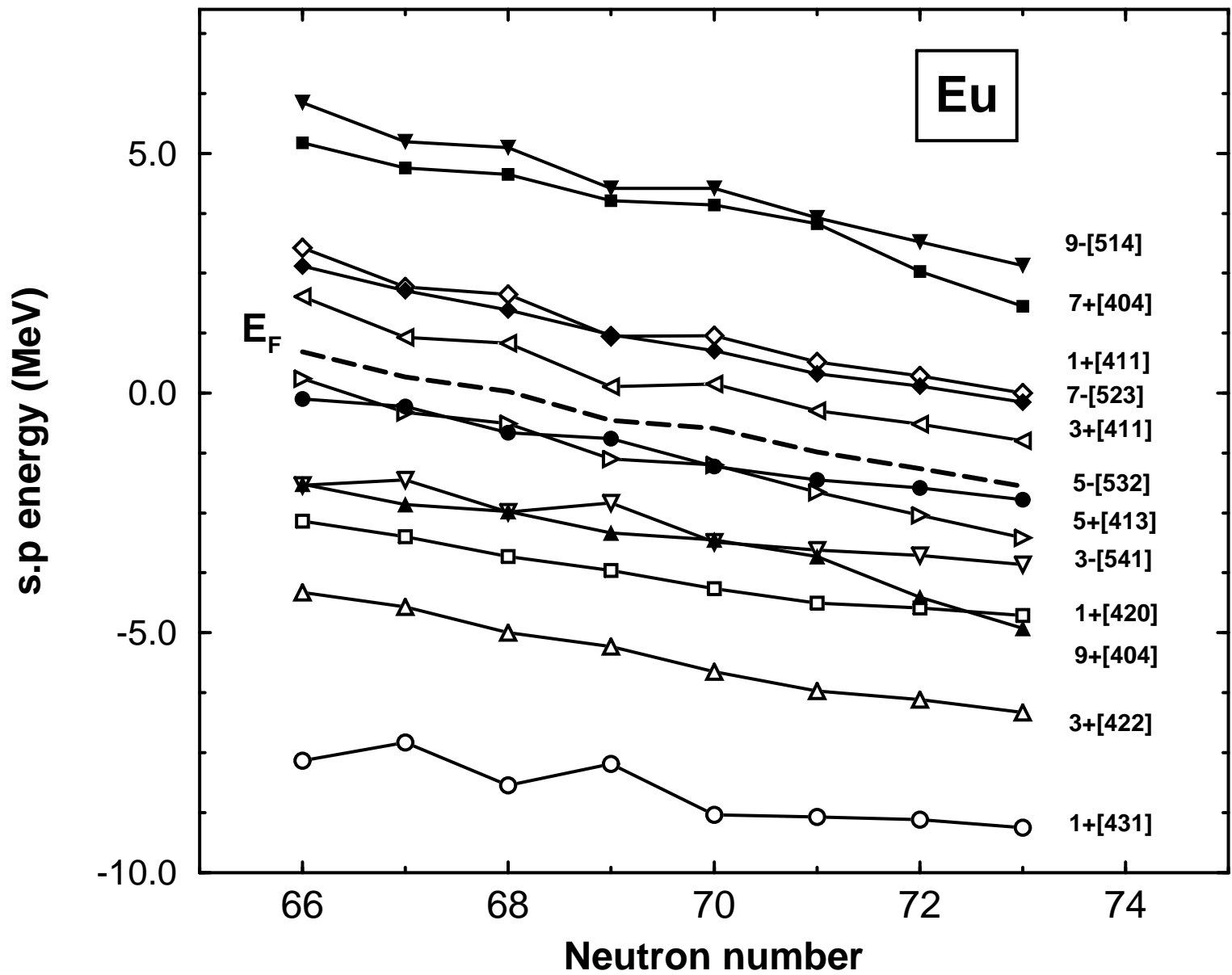
\end{document}